\documentclass[twocolumn,secnumarabic,amssymb, nobibnotes, aps, prd]{revtex4}

\usepackage{graphicx}

\begin{document}

\title{Spacetime constraints on accreting black holes}%

\author{David Garofalo} \email[]{David.A.Garofalo@jpl.nasa.gov}
\affiliation{Jet Propulsion Laboratory, California Institute of
Technology, Pasadena CA 91109} %

\begin{abstract}
We study the spin dependence of accretion onto rotating Kerr black
holes using analytic techniques.  In its linear regime, angular
momentum transport in MHD turbulent accretion flow involves the
generation of radial magnetic field connecting plasma in a
differentially rotating flow.  We take a first principles approach,
highlighting the constraint that limits the generation and
amplification of radial magnetic fields, stemming from the transfer of
energy from mechanical to magnetic form.  Because the energy
transferred in magnetic form is ultimately constrained by
gravitational potential energy or Killing energy, the spin-dependence
of the latter allows us to derive spin-dependent constraints on the
success of the accreting plasma to expel its angular momentum and
accrete.  We find an inverse relationship between this ability and
black hole spin.  If this radial magnetic field generation forms the
basis for angular momentum transfer in accretion flows, accretion
rates involving Kerr black holes are expected to be lower as the black
hole spin increases in the prograde sense.
\end{abstract}

\maketitle

\section{Introduction}

Magnetized accretion onto rotating black holes forms the basis for
models of X-ray binaries (XRBs), active galactic nuclei (AGN) and
their smaller counterparts, microquasars.  Balbus \& Hawley (1991)
showed how the magnetorotational instability (MRI) - discovered
independently by Velikhov and Chandrasekhar - could lead to angular
momentum transport in differentially rotating MHD flow (Velikhov,
1959; Chandrasekhar, 1961), such as in magnetized accretion onto
rotating black holes.  For a sufficiently small ratio of magnetic
pressure to gas pressure, the highly conducting flow advects the
magnetic field with it in a process described as flux-freezing.  This
means that the magnetic field is subject to being deformed and
stretched as determined by the gas motion, producing magnetic
connections between spatially distinct regions.  If gas at an inner
accretion disk radius, $r_{in}$, is coupled magnetically to gas at an
outer radius, $r_{out}$, the magnetic field transports angular
momentum from $r_{in}$ to $r_{out}$, decreasing the angular momentum
of gas at $r_{in}$ and increasing the angular momentum of gas at
$r_{out}$.  As a consequence of this, gas at $r_{in}$, which loses
angular momentum, spirals inward to smaller radial values and larger
circular velocities.  Gas at $r_{out}$, instead, migrates to larger
radial values compared to $r_{out}$, compatible with the angular
momentum acquired, and settles in circular orbits with lower circular
velocities.  As long as inward spiraling gas continues to be
magnetically coupled to outward migrating gas, angular momentum is
transferred at a greater rate as the coupling distance increases.  Due
to the energetically less favorable configurations, larger coupling
distances are limited, and eventually suffer magnetic reconnection.
This study highlights the constraints from energy conservation on
coupling distance, ultimately suggesting that general relativity
imposes a coupling distance constraint that depends monotonically on
the spin of the black hole.  The monotonic nature of this dependence,
suggests that jets and outflows produced in magnetized black hole
accretion flows, may bear the signature of black hole spin.  Section
\ref{Work} describes the study and presents the results.  Section
\ref{Discussion} concludes.

\section{Spin-dependence of accretion flows}
\label{Work}

Under the assumption that the MRI is the basic mechanism behind
angular momentum transport, we measure the ability of a thin accretion
disk to expel its angular momentum and accrete, by determining the
amount of work done to extract the angular momentum that exists over a
fixed proper distance between $r_{in}$ and $r_{out}$. For a fixed mass
density flow between $r_{out}$ and $r_{in}$, the angular momentum
difference between $r_{out}$ and $r_{in}$ must be deposited outward of
$r_{out}$. We assume that the ability to extract that angular momentum
difference is the same for all black hole spins and show how this
assumption becomes increasingly problematic at higher prograde spin.
From conservation laws in the Kerr metric, the angular momentum
difference that must be extracted to produce a given accretion rate,
increases for fixed proper distance as the black hole spin increases
in the prograde sense.  A work-kinetic energy argument applied to
angular momentum extraction follows, motivating a spin-dependent
constraint.  We begin by introducing two conserved quantities in the
Kerr spacetime that originate from time-translation and azimuthal
angle-invariance that are identified as energy and angular momentum
per unit mass for circular geodesic orbits in the equatorial plane of
a rotating black hole.  From the Kerr metric we have

\begin{equation}
E = -\epsilon_{\nu}p^{\nu}= (1-\frac{2M}{r})\dot{t} +
\frac{2Ma}{r}\dot{\phi}
\end{equation}

\begin{equation}
L = \psi_{\nu}p^{\nu} = -\frac{2Ma}{r}\dot{t} +
\frac{(r^{2}+a^{2})^{2}-\Delta a^{2}}{r^{2}}\dot{\phi}
\end{equation}

where $\epsilon$ and $\psi$ are the two spacetime Killing vectors for
Boyer-Lindquist coordinates, $p^{\nu}$ is the 4-momentum, dot implies
differentiation with respect to proper time, $M$ and $a$ are the mass
and spin parameters, $r$ is the radial coordinate, $t$ is the time
coordinate, $\phi$ is the azimuthal angle coordinate, and $\Delta =
r^2 - 2Mr + a^2$.  Replacing $\dot{t}$ and $\dot{\phi}$ in terms of
metric components, one obtains the following explicit forms for E and
L (Bardeen et al, 1972).

\begin{equation}
E = \frac{r^{3/2}-2Mr^{1/2}\pm aM^{1/2}} {r^{3/4}(r^{3/2}-3Mr^{1/2}\pm
2aM^{1/2})^{1/2}}
\end{equation}

\begin{equation}
L = \pm M^{1/2}\frac{r^{2}\mp 2aM^{1/2}r^{1/2}+a^{2}}
{r^{3/4}(r^{3/2}-3Mr^{1/2}\pm2aM^{1/2})^{1/2}}
\end{equation}

with upper sign for prograde orbits and lower sign for retrograde orbits.

\begin{figure*}
 \includegraphics{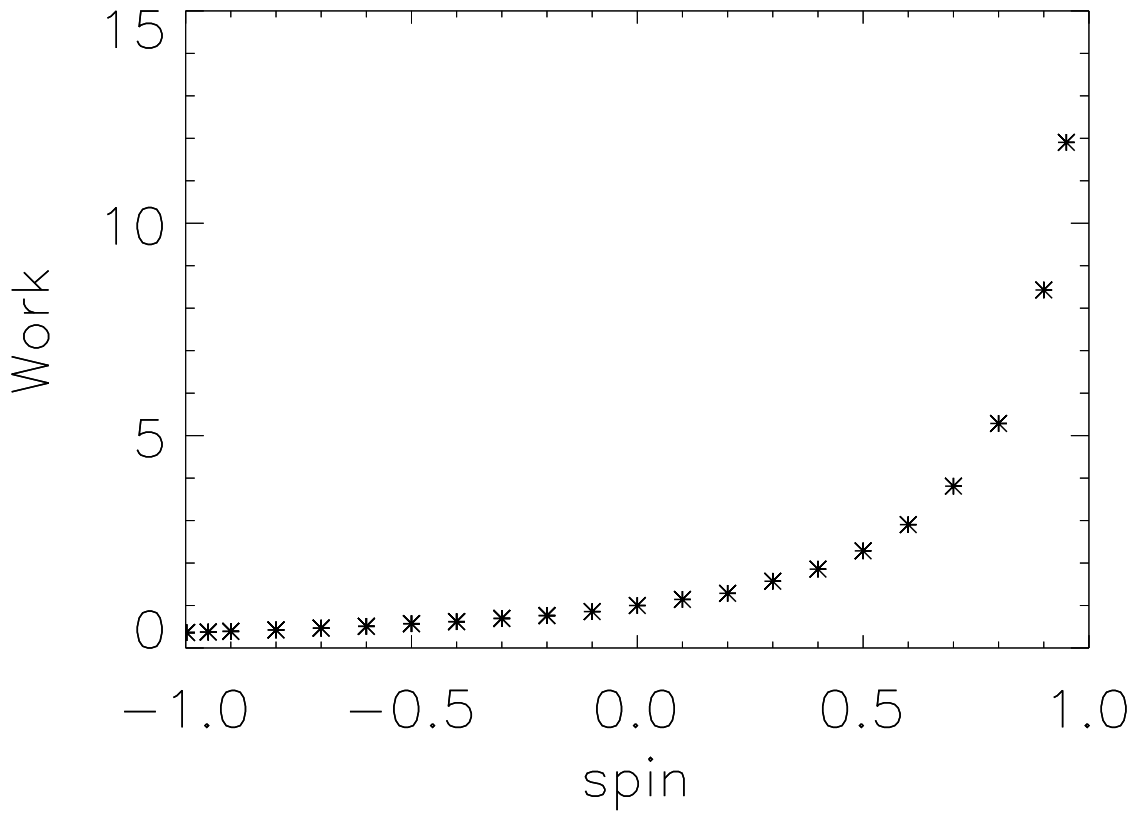}
 \caption{\label{Work-spin}Work vs. spin illustrating the difficulty
  in extracting the angular momentum at higher prograde spin.
  Negative values indicate retrograde accretion while positive ones
  represent prograde accretion.}
 \end{figure*}

As mentioned, we choose a fixed proper distance, R, from the
marginally stable circular orbit and determine the energy difference
between the marginally stable orbit and the radial location for which
the proper distance from the marginally stable circular orbit is R.
We show the spin dependence of this energy difference in figure
\ref{Work-spin} and refer to it as work resulting from the difference
in kinetic energy of the gas parcel between $r_{in}$ and $r_{out}$.
We conclude that if the accretion rate is to remain spin-independent,
the gas must lose greater energy per mass density, per unit coordinate
time, as the spin becomes more prograde.  

Given that it is the MRI behind the transfer of angular momentum and
energy between $r_{in}$ and $r_{out}$, we motivate the existence of a
spin-dependence in the energy transferred from mechanical form in the
gas, to magnetic form in the following way.  We imagine that two gas
parcels, initially adjacent and connected via magnetic field (at some
radial position that is intermediate between $r_{out}$ and $r_{in}$),
begin to slowly migrate apart.  If the two initially adjacent gas
parcels become separated by a proper distance R, as assumed, and their
radial velocities are small, the energy transferred from kinetic to
magnetic form will roughly be equal to that of figure \ref{Work-spin}
for the specific spin.  Therefore, in order for accretion rates to be
spin-independent, the energy in the poloidal component of the magnetic
field, must increase with increase in prograde spin.  But, magnetic
reconnection is progressively more likely to occur before the proper
coupling distance becomes R, as the spin increases in the prograde
direction.  The overall conclusion, thus, is that lower accretion
rates are energetically preferred as the spin increases in the
prograde direction.  One might be tempted to consider the spin
dependence of the ability to transport angular momentum by focusing on
the work required to extract a fixed amount of angular momentum.  In
other words, instead of approaching the question by determining the
work done over a fixed proper distance, consider the work required to
extract a fixed angular momentum difference.  Although we claim this
not to be the appropriate approach, it is perhaps noteworthy that if
one were to calculate work vs. spin this way, the trend would be the
same (i.e. that greater work is required to extract that fixed angular
momentum difference as the spin increases in the prograde direction).
Either way, the direct connection assumed here between the energy
difference over a fixed proper distance and the likelihood of magnetic
reconnection, is speculative.  It could be that magnetic reconnection
is related to aspects of the Kerr geometry in other non-trivial ways
that are not addressed here.  Also, since the constraints discussed
here depend on the spin parameter of the black hole, they are purely
relativistic, so they produce no Newtonian counterpart to this study.

\section{Conclusion}
\label{Discussion}

This work highlights features of spacetime that appear to be
intrinsically suited to influence the character of MHD accretion
flows.  Because MHD accretion flows in AGN are linked to outflows that
influence the galactic and intergalactic medium (Kormendy \&
Richstone, 1995; Magorrian et al. 1998; Marconi \& Hunt, 2003;
Gebhardt et al. 2000; Ferrarese \& Merritt, 2000; Tremaine et al,
2002), black hole spin may produce signatures or even influence the
evolution of galaxies (Garofalo, 2009).  If the MRI forms the
foundation for angular momentum transport in magnetized accretion
flows, and, given that its modus operandi is based on magnetically
connected spatially separated regions, we have derived constraints
that influence the spatial connectivity of such regions.  Our
suggestion is that the spin dependence of the conservation laws may be
reflected in the flow in terms of dynamical constraints.  Our analysis
in section \ref{Work} points to the need for greater conversion of
energy to magnetic form in order for the accretion rate to be
insensitive to black hole spin.  From an energetic viewpoint, then,
this suggests that the magnetized flow will be subjected to a spin
dependence involving lower accretion rates at higher prograde spin.
If MHD turbulence is non-local (e.g. Guan et al., 2009), our analysis
applies.  We note that accretion rates in GRMHD simulations of
adiabatic black hole accretion flows, do in fact decrease as the spin
increases in the prograde direction (De Villiers, 2003; McKinney,
2004).

\section{Acknowledgments}

 The research described in this paper was carried out at the Jet
Propulsion Laboratory, California Institute of Technology, under a
contract with the National Aeronautics and Space Administration.
D.G. is supported by the NASA Postdoctoral Program at NASA JPL
administered by Oak Ridge Associated Universities through contract
with NASA.  


\section{References}

\noindent Balbus, S.A., \& Hawley, J.F., 1991, ApJ, 376, 214

\noindent Velikhov, F., 1959, Soviet Phys-JETP, 36, 1398

\noindent Chandrasekhar, 1961, Hydrodynamic and Hydromagnetic
Stability, Oxford University Press

\noindent Bardeen, J.M., Press, W.H., Teukolsky, S.S., 1972, ApJ, 178,347

\noindent Kormendy, J., \& Richstone, D., 1995, ARA\&A, 33, 581

\noindent Magorrian, J., et al. 1998, AJ, 115, 2285

\noindent Marconi, A., \& Hunt, L.K., 2003, ApJ, 589, L21

\noindent Gebhardt, K., et al. 2000, ApJ, 539, L13 

\noindent Ferrarese, L., \& Merritt, D., 2000, ApJ, 539, L9

\noindent Tremaine, S., et al 2002, ApJ, 574, 740

\noindent Garofalo, D., 2009, ApJL, in press

\noindent Guan, X. et al., 2009, ApJ, 694, 1010.

\noindent De Villiers, J.P. et al., 2003, ApJ, 599, 1238.

\noindent McKinney, J.C., Gammie, C.F., 2004, ApJ, 611, 977

\end{document}